\documentclass[sigconf,nonacm]{acmart}

\usepackage{booktabs}
\usepackage{array}
\usepackage{graphicx}
\usepackage{makecell}
\usepackage{xcolor}

\usepackage{amssymb}
\usepackage{subcaption}
\usepackage{tikz}
\usepackage{placeins}

\usetikzlibrary{shapes.geometric, arrows.meta, positioning, fit}

\graphicspath{{figs/}}

\title{AgentClick: A Skill-Based Human-in-the-Loop\\Review Layer for Terminal AI Agents}

\author{Haomin Zhuang}\authornote{Equal Contribution}
\affiliation{\institution{University of Notre Dame} \country{United States}}

\author{Hanwen Xing}\authornotemark[1]
\affiliation{\institution{University of Southern California} \country{United States}}

\author{Xiangliang Zhang}
\affiliation{\institution{University of Notre Dame} \country{United States}}

\begin{document}

\begin{abstract}
Recent autonomous AI agents such as Codex, and Claude Code have made it increasingly practical for users to delegate complex tasks, including writing emails, executing code, issuing shell commands, and carrying out multi-step plans. However, despite these capabilities, human--agent interaction still largely happens through terminal interfaces or remote text-based channels such as Discord. These interaction modes are often inefficient and unfriendly: long text outputs are difficult to read and review, proposed actions lack clear structure and visual context, and users must express feedback by typing detailed corrections, which is cumbersome and often discourages effective collaboration. As a result, non-expert users in particular face a high barrier to working productively with agents.

To address this gap, we present \textbf{AgentClick}, an interactive review layer for terminal-based agents. AgentClick is implemented as a localhost npm server paired with a skill-based plugin that connects the running agent to a browser interface, allowing users to supervise and collaborate with agents through a structured web UI rather than raw terminal text alone. The system supports a range of human-in-the-loop workflows, including email drafting and revision, plan review and modification, memory management, trajectory inspection and visualization, and error localization during agent execution. It also turns code generation and execution into a reviewable process, enabling users to inspect and intervene before consequential actions are taken. In addition, AgentClick supports persistent preference capture through editable memory and remote access over HTTP, allowing users to review agents running on servers from their personal devices. Our goal is to lower the barrier for non-expert users and improve the efficiency and quality of human--agent co-work. The code is available at \url{https://github.com/agentlayer-io/AgentClick/tree/main}.
\end{abstract}
\maketitle
\vspace{-0.1in}

\vspace{-0.1in}
\section{Introduction}
\label{sec:intro}
\begin{figure}
    \centering
    \includegraphics[width=1\linewidth]{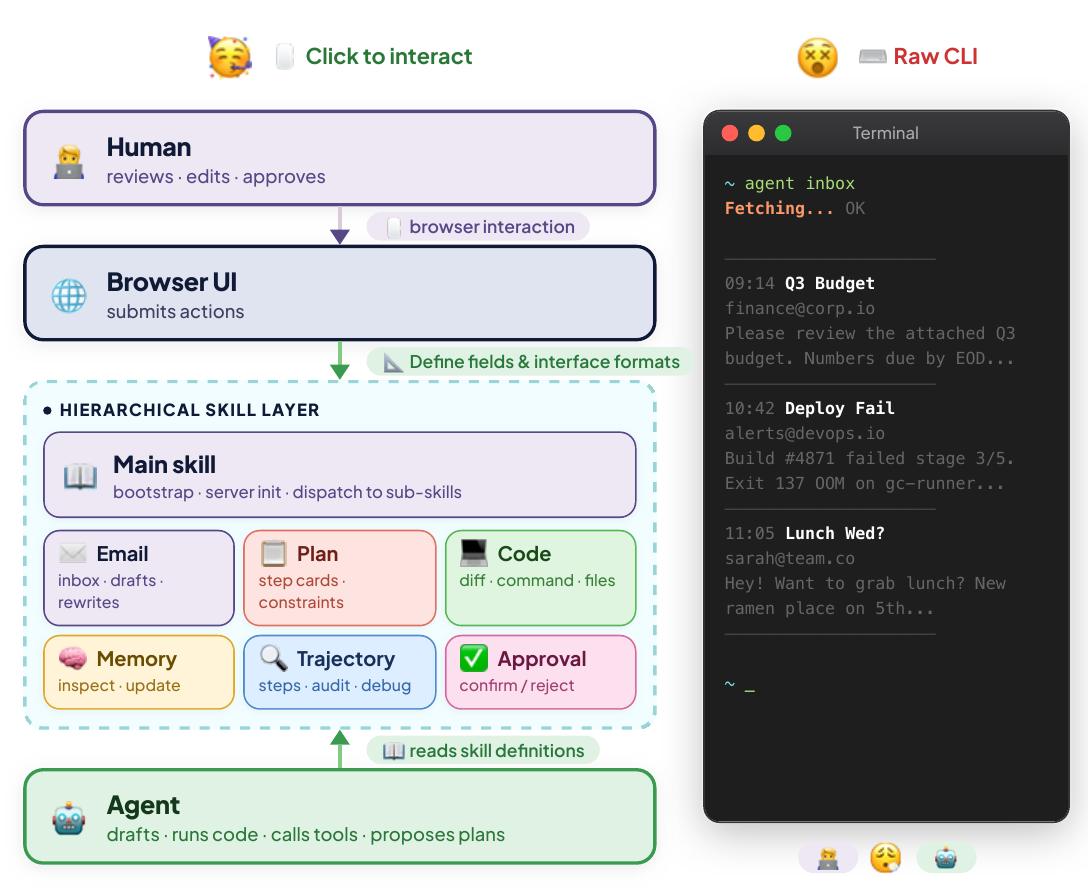} 
    \caption{ Demo of email review via AgentClick vs.   terminal }
    \label{fig:overview}
    \vspace{-0.25in}
\end{figure}
Agentic AI systems are increasingly capable of carrying out long-horizon tasks such as triaging inboxes, executing code pipelines, and orchestrating multi-step workflows~\cite{guo2024large,gronauer2022multi}. Recent systems such as Codex~\cite{openai2025codexcli}, OpenClaw~\cite{steinberger2025openclaw}, and Claude Code~\cite{anthropic2024claudecode} make it practical for users to delegate increasingly consequential work to agents, including writing emails, planning actions, editing files, running shell commands, and interacting with external tools. This shift creates new opportunities for productivity, but also raises a central interaction challenge: how can humans effectively collaborate with autonomous agents while retaining meaningful visibility, control, and opportunities for correction~\cite{mozannar2025magenticui}?
At the current stage,
interaction with such agents still happens primarily through terminal interfaces or remote text-based channels such as Discord. These modes are poorly suited to human--agent co-work for three reasons. First, terminal output interleaves reasoning traces, tool logs, and proposed actions in a single stream, making it difficult to identify what requires review. Second, feedback is cumbersome: users must type free-form corrections even for small local edits. Last,  consequential actions---sending email, executing code, committing to a plan---are presented as opaque events or binary prompts, leaving little room for targeted inspection or modification. These barriers are especially pronounced for non-expert users, and become more acute when agents run on remote or headless infrastructure, where the terminal is not only a poor collaboration medium but often an inaccessible one.
We present \textbf{AgentClick}, an interactive review layer for terminal-based agents that improves human--agent collaboration rather than merely gating execution. AgentClick runs as a localhost npm server paired with a skill-based plugin, exposing a browser-based interface in place of raw terminal text. It surfaces structured review UIs tailored to different action types: email drafting, plan editing, memory management, trajectory visualization, and code inspection. Users interact at the level of the artifact being produced rather than at the level of terminal output. AgentClick also supports persistent preference capture, so user edits feed into agent memory for future sessions, and serves its interface over HTTP for remote access from personal devices. Figure~\ref{fig:overview} contrasts a terminal-only session with the same session in AgentClick's Email Review UI.

\vspace{-0.1in}
\section{System Design}
\subsection{Human--Browser--Backend--Agent Architecture} \vspace{-0.1in}
AgentClick is built around four components: \emph{human}, \emph{browser}, \emph{backend}, and \emph{agent}. The \emph{agent} executes tasks autonomously—drafting emails, proposing plans, running code, and calling tools—but rather than acting immediately, it submits proposals to the backend and waits for a review outcome. The \emph{backend} acts as the coordination layer: it receives proposals from the agent, maintains session state and user preferences, serves the browser interface over HTTP, records user edits, and returns approval or correction signals to the agent. The \emph{browser} exposes task-specific views tailored to different action types—email review, plan editing, memory inspection, trajectory visualization, and code review—giving the user structured access to agent proposals at the artifact level rather than the terminal level. The \emph{human} reviews, edits, and approves these proposals through the browser, with corrections optionally persisted to agent memory for future sessions. 

\vspace{-0.15in}
\subsection{Interactive Review Flow}\vspace{-0.1in}
When the \emph{agent} reaches a consequential step, it submits a proposal to the \emph{backend}, which creates a review session and surfaces it in the \emph{browser} UI. Depending on the artifact type, the \emph{human} user may approve, directly edit, delete content, adjust constraints, or request a targeted rewrite; the result is returned to the agent, which incorporates the feedback and continues. This goes beyond binary accept-or-reject: users can intervene at the level of the artifact, which is particularly useful when output is mostly correct but requires localized changes.
The same flow supports monitoring and debugging. Because intermediate artifacts are surfaced before final execution, users can track agent progress, locate errors early, and intervene before a problematic action is committed.

\vspace{-0.15in}
\subsection{Hierarchical Skill-Based Integration}\vspace{-0.1in}
AgentClick connects the backend and the agent through a hierarchical skill design. Rather than requiring a dedicated SDK or direct code-level integration, the agent consumes a structured set of Markdown skills that define how to launch AgentClick, communicate with the backend, and invoke task-specific review workflows. This approach has two key advantages. First, skills are lightweight: they require no instrumentation of the agent codebase and add no runtime dependencies. Second, they are portable: any agent that can read and follow Markdown instructions—such as Claude Code or OpenClaw—can adopt AgentClick without modification. In this sense, skills act as the glue layer between agent execution and browser-based human review~\cite{xing2026recipes}.

At the top level, a \emph{main skill} handles shared integration logic and serves as both a bootstrap mechanism and a dispatcher. It instructs the agent how to initialize or connect to the local AgentClick server, establishes the common protocol for submitting proposals and receiving review results, and routes the agent to the appropriate sub-skill based on task intent—without requiring the agent to implement each review mode from scratch.

Below this, AgentClick defines a set of \emph{sub-skills} for concrete task
scenarios such as email, planning, trajectory inspection, code review, memory
management, and generic approval. Each sub-skill specifies the task-specific
payload schema, expected review actions, and result handling conventions for its
domain. For example, the email sub-skill defines how inbox items, drafts, and
paragraph-level rewrites are represented; the planning sub-skill defines how
steps, constraints, and edits are exchanged; and the trajectory sub-skill
defines how intermediate execution traces are surfaced for inspection. This
hierarchical design separates shared coordination logic from domain-specific
review behavior, making the system easier to extend with new review modes over
time.

\begin{figure}[t]
  \centering
  \includegraphics[width=\columnwidth]{figs/email.png}
  \vspace{-0.2in}
  \caption{%
  \textbf{Terminal vs.\ AgentClick Browser UI for email review.}
  In the terminal (top), agent-drafted emails are hard to review and edit. AgentClick (bottom) renders the full inbox,
  message content, and editable reply drafts, with execution gated on
  explicit user confirmation.
}
  \label{fig:email_comparison}
  \vspace{-0.3in}
\end{figure}

\vspace{-0.15in}
\subsection{Task-Specific Review Interfaces}\vspace{-0.1in}

A key design principle of AgentClick is that different agent outputs require
different review affordances. Instead of presenting all agent proposals in a
single generic interface, AgentClick provides task-specific browser UIs that
match the structure of the artifact.

\textbf{Email Review.} Although agents can already connect to Gmail via tool calls, email interaction through a terminal remains impractical: users have no convenient way to browse and select specific messages, raw text rendering makes threads hard to read, and modifying a generated reply requires retyping an entire instruction even for a small change. AgentClick's Email UI addresses these limitations by combining inbox browsing, full message rendering, and reply drafting in a single interface (refer to Fig.~\ref{fig:email_comparison}). Users can select a specific email, request a draft only when needed, and edit the reply at the paragraph level with per-section rewrite, edit, and delete controls. The agent monitors user actions in real time and can update drafts accordingly. Final sending is gated on explicit user confirmation. Figure~\ref{fig:email_comparison} illustrates the contrast between terminal-based email interaction and the AgentClick Email UI.

\textbf{Plan Review.} Terminal-based planning has several practical problems. Agent reasoning and plan steps are interleaved in a single output stream, forcing users to scroll back to reconstruct the full plan. Long plans are tedious to read as raw text. And when the agent pauses for user input mid-plan, users must make choices without visibility into subsequent steps—while any conversational exchange at that point risks disrupting the agent's ability to surface later decision points correctly. AgentClick's Plan UI addresses these issues by presenting the entire proposed workflow upfront as structured step cards before execution begins. Each card is typed and annotated—distinguishing tool calls, file operations, code execution, and other action categories—so users can scan the plan at a glance and direct attention to steps that warrant closer review. Individual steps can be edited in place: users may modify descriptions, adjust constraints, reorder steps, or remove unnecessary actions. This structured representation makes long-horizon plans both more legible and more precisely correctable than their terminal equivalents.

\begin{figure}[!htbp]
    \centering
    \includegraphics[width=1\linewidth]{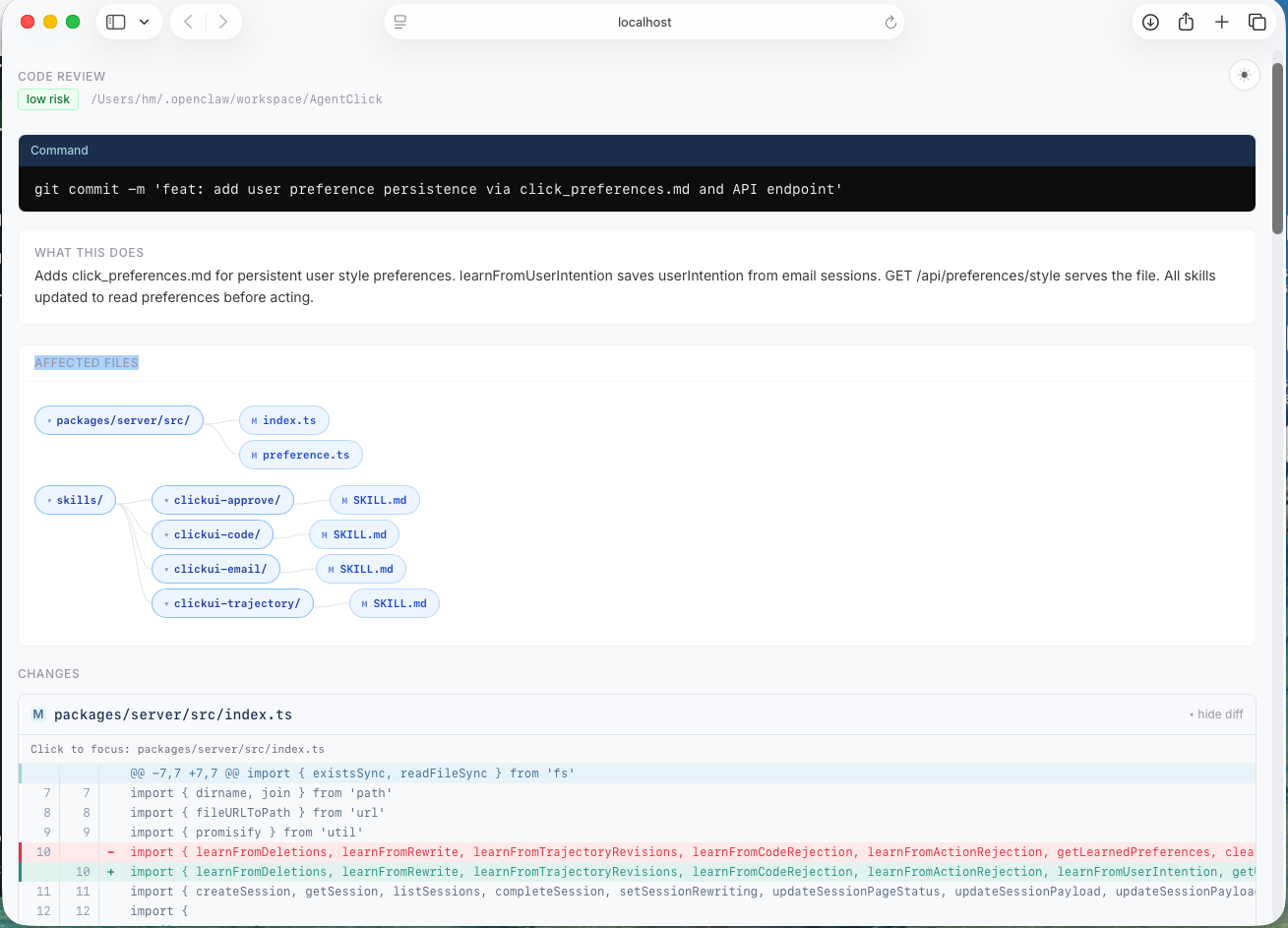}
    \caption{Click UI for code review.}
    \label{fig:code}
\end{figure}

\textbf{Code Review.} Although agents already output diffs for proposed code changes, terminal-based review has practical limitations: large changesets are displayed as a flat stream of additions and deletions that is difficult to parse, and diff output lacks surrounding context, making it hard to understand whether a change is correct without opening the file separately. AgentClick's Code UI addresses this by presenting each proposed change as a structured review artifact (refer to Fig.~\ref{fig:code}). Users see the proposed command alongside a natural-language explanation, a list of affected files, and a unified diff—but can also expand any file to view its full content with the change in context. Rather than accepting or rejecting the entire changeset, users can approve individual hunks, annotate specific lines with targeted instructions, or request a partial rewrite of selected sections. This turns code execution from an opaque terminal event into a reviewable, precisely correctable process.

\textbf{Memory Management.} Memory handling is a persistent pain point in current agent systems. Agents typically abstract over memory operations entirely, giving users no visibility into what is being stored, retrieved, or discarded. The problem is especially acute during context compaction: when a session grows long, the agent automatically summarizes and clears the conversation, but the resulting summary often omits content the user actually needs, and any memory that should persist must be manually reloaded in the next session by locating the relevant file in the project tree. AgentClick's Memory UI addresses this by making memory state explicit and controllable at every stage. Users can inspect all currently loaded memory entries, review proposed updates before they are committed, and guide how session memory is constructed during compaction. When compaction occurs, AgentClick generates a structured memory summary and surfaces it for user review and editing before it is saved (refer to Fig.~\ref{fig:memory} in Appendix). All available memory files are listed in a single panel, allowing users to load or unload entries with a single click rather than navigating the file system. This gives users direct control over what the agent knows, what it forgets, and what carries forward across sessions.

\textbf{Trajectory Inspection.} In current agent systems, execution trajectories are hidden by default: users see only the final output, with no visibility into the sequence of tool calls, retries, and intermediate decisions that produced it. This makes it difficult to diagnose unexpected behavior—such as excessive token consumption—even after a task completes successfully. AgentClick's Trajectory UI makes the full execution trace visible and structured (refer to Fig.~\ref{fig:traj} in Appendix). Tool calls, failed attempts, and error recoveries are highlighted distinctly, allowing users to follow the agent's decision path at a glance and pinpoint where inefficient or undesirable behavior occurred. Users can also annotate specific steps with guidance, which is saved to memory and used to inform the agent's behavior on similar tasks in future sessions. This turns trajectory inspection from a post-hoc debugging tool into an active channel for human oversight and iterative agent improvement.

\vspace{-0.1in}
\subsection{Remote and Cross-Device Review}
\vspace{-0.1in}
Although AgentClick starts as a local server, an important deployment setting is agents running on remote or headless infrastructure such as cloud VMs, shared servers, or containers. In these cases, users often interact with agents indirectly through text-based channels such as Discord, with no convenient access to the underlying terminal. AgentClick addresses this by exposing the review layer over HTTP, so the same browser-based interfaces are accessible remotely from any device. For example, a user running an OpenClaw agent on a Mac Mini can receive an approval prompt in Discord, open the AgentClick URL on their phone, review the full email draft or plan in the browser, make edits, and confirm—without ever touching the terminal. Rather than parsing long textual updates or typing corrections through chat, users can inspect drafts, adjust plans, monitor trajectories, and approve actions through a structured web interface on their own device.
\vspace{-0.1in}
\subsection{Preference Memory and Adaptation}
\vspace{-0.1in}
To reduce repeated correction effort across sessions, AgentClick captures user preferences from review interactions. When users edit or delete parts of an artifact and provide an explicit reason, the system records this feedback in a structured \texttt{MEMORY.md} file that the agent incorporates in future runs. This is especially useful for recurring tasks such as email writing, where similar stylistic corrections would otherwise be applied repeatedly. Over time, the review layer functions not only as an approval interface for current actions, but also as a lightweight mechanism for accumulating user preferences and improving future human--agent co-work.

\vspace{-0.05in}
\section{Evaluation}
\vspace{-0.05in}

\paragraph{Evaluation Goals}We evaluate AgentClick through the lens of human--agent collaboration rather than autonomous task completion in isolation. Our goal is to assess whether AgentClick fills the interaction gaps identified in Section~\ref{sec:intro}: limited visibility into proposed actions, cumbersome localized correction, and weak support for remote or cross-device oversight. We focus on whether the system enables users to communicate intent more effectively, intervene on structured artifacts before execution, and supervise agents in deployment settings where terminal-only interaction is inconvenient or unavailable.
We exercise these three capabilities through representative task walkthroughs: plan review and constraint injection before execution, targeted artifact rewrite for better alignment with user intent, and remote review of an agent running on headless infrastructure. 
In this demo paper, we treat these walkthroughs as capability-oriented illustrations rather than a formal controlled user study.
We release the full codebase and demo scripts (and a video) to support reproducibility and enable others to verify and extend the showcased workflows (\url{https://github.com/agentlayer-io/AgentClick/tree/main}).
\vspace{-0.2in}
\subsection{Task Walkthroughs}
\begin{figure}[!htbp!]
    \centering
    \begin{subfigure}{0.47\textwidth}
        \centering
        \includegraphics[width=\linewidth]{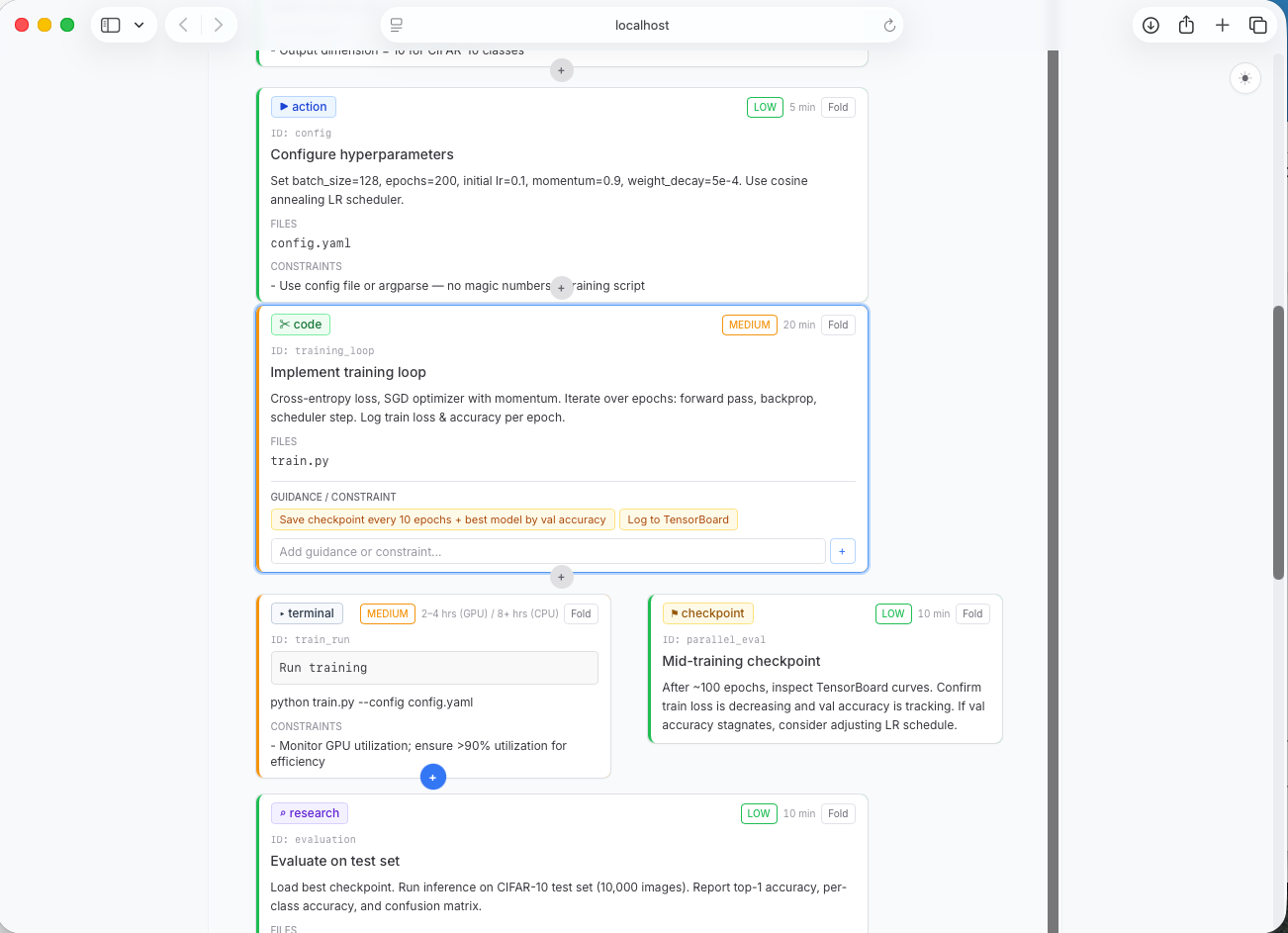}
        \caption{Planning Browser UI}
    \end{subfigure}
    \hfill
    \begin{subfigure}{0.47\textwidth}
        \centering
        \includegraphics[width=\linewidth]{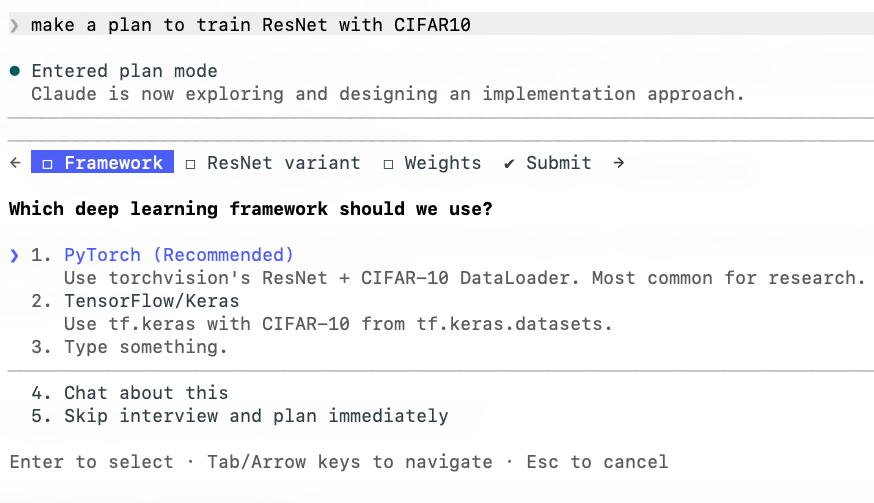}
        \caption{Planning TUI}
    \end{subfigure}
    \caption{ Planning Click UI vs. TUI: Global visibility enables more informed framework selection}
    \label{fig:planning}
\end{figure}

\vspace{-0.1in}
\textbf{Task 1 --- Plan review and targeted constraint injection.}
As shown in Fig.~\ref{fig:planning}, Claude Code was asked to generate a plan for training ResNet-18 on CIFAR-10.
In the terminal, the agent surfaced the plan as a multiple-choice prompt
(``Which framework? 1.~PyTorch 2.~TensorFlow 3.~Type something''), offering
no visibility into downstream steps at the time of choosing. In AgentClick's
Plan UI, the same plan was presented as structured step cards showing the full
pipeline, including framework configuration, training loop implementation,
execution, checkpointing, and final evaluation. The user then injected two
targeted constraints before execution: for the training loop step, ``Save
checkpoint every 10 epochs + best model by val accuracy,'' and for the
execution step, ``Monitor GPU utilization; ensure $>$90\% utilization for
efficiency.'' The agent subsequently executed the modified plan with these
constraints incorporated. This walkthrough demonstrates how AgentClick allows
users to express intent at the level of the plan artifact itself, rather than
requiring them to anticipate all constraints upfront in a terminal exchange or
interrupt the workflow later after execution has begun.

\begin{figure}[!htbp]
    \centering
    \includegraphics[width=0.95\linewidth]{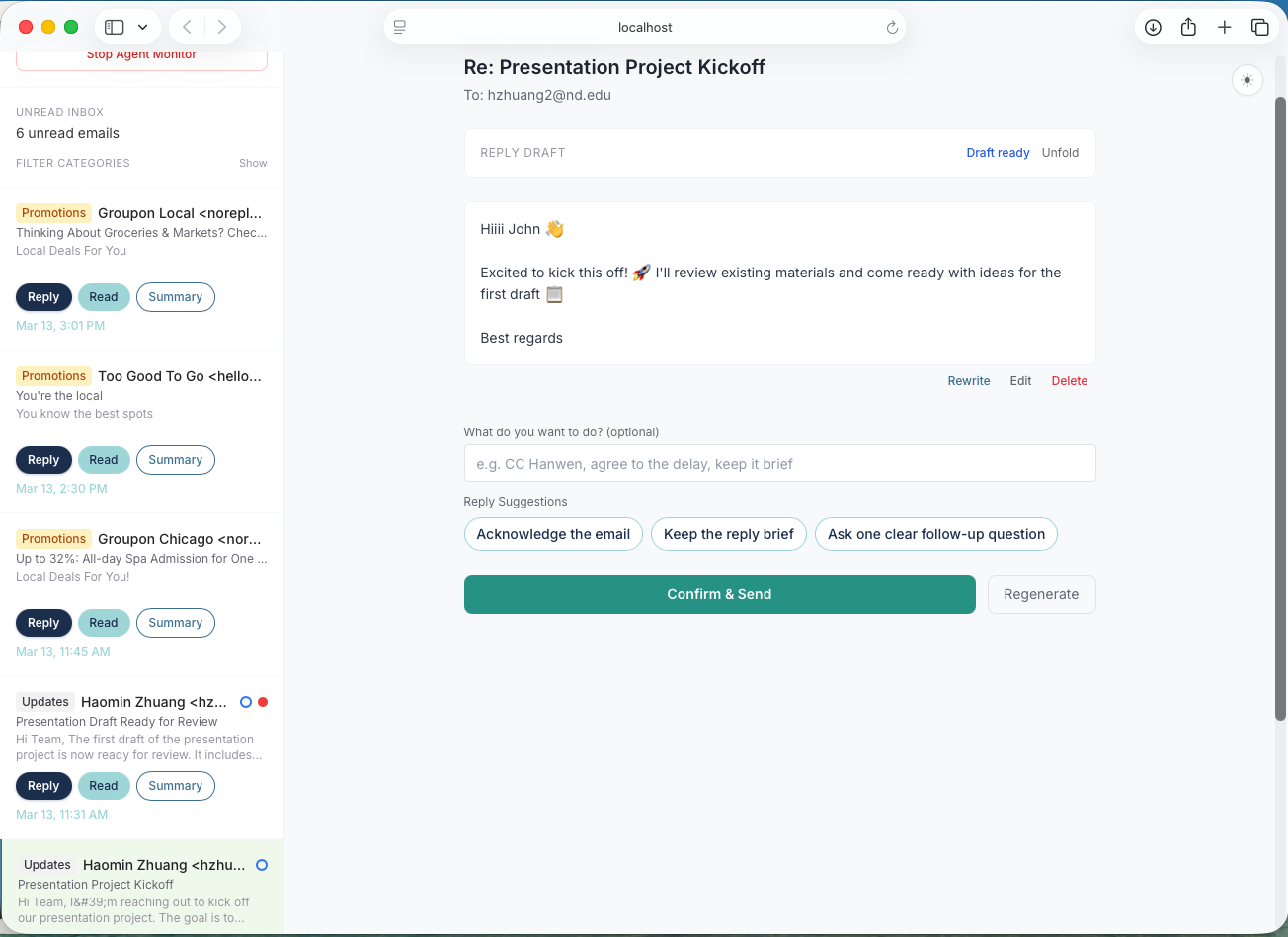}
    \caption{The email reply learns from preference then generate reply with Emoji by default.}
    \label{fig:email_emoji}
\end{figure}

\vspace{-0.02in}
\textbf{Task 2 --- Email rewrite for intent preference.}
Claude Code was asked to reply to an email about a presentation draft. The
agent's initial reply was written in formal English and did not match the
user's preferred tone. In AgentClick's Email UI, the user selected a paragraph
for rewriting and provided an explicit reason requesting that the response use
emoji and adopt a lighter style. The revised draft was then confirmed and sent.
As shown in Fig.~\ref{fig:email_emoji}, when the agent generated a reply for the next email, it
automatically adopted the updated stylistic preference without requiring the
user to restate the instruction. This walkthrough highlights a central goal of
AgentClick: improving the agent's ability to infer and operationalize human
intent through structured review. Rather than treating correction
as a one-off override, the system keeps the agent attached to the live review
context, allowing user-provided rewrite reasons to shape subsequent generation
in the interaction loop.

\vspace{-0.02in}
\textbf{Task 3 --- Remote review from a mobile device.}
To evaluate AgentClick in a remote deployment setting, we ran an agent under
OpenClaw on a server and exposed AgentClick through a Cloudflare tunnel. The
resulting URL was opened in a mobile browser, where the user reviewed an email
draft in the AgentClick UI, edited one paragraph, and confirmed the result.
The server-side agent then received the approval and dispatched the email.
This walkthrough demonstrates that AgentClick extends human-in-the-loop review
to settings where the agent and the user are physically separated and the
terminal is not the primary interface. In contrast to terminal-only workflows
or chat-based relays such as Discord, the user can inspect and modify the
artifact directly from a personal device through a structured browser
interface.
\vspace{-0.1in}
\subsection{Capability Coverage}
\vspace{-0.1in}
Table~\ref{tab:capabilities} summarizes the capabilities exercised by the three walkthroughs. Task~1 shows that AgentClick supports pre-execution intervention on long-horizon plans, allowing users to refine downstream behavior before expensive execution begins. Task~2 shows that targeted rewriting with explicit reasons improves alignment between agent output and user intent, reducing the need for repeated restatement. Task~3 shows that the same review model remains usable when agents run on remote or headless infrastructure, supporting cross-device oversight without terminal access. Across all three tasks, the key benefit is not simply that AgentClick inserts a pause before execution, but that it gives users a more effective way to communicate intent and make localized corrections at the level of the artifact under consideration—whether a plan, an email draft, or a remote approval request.

\begin{table}[t]
  \centering
  \caption{Capabilities exercised in the evaluation tasks } 
  \vspace{-0.1in}
  \label{tab:capabilities}
  \scalebox{0.7}{\begin{tabular}{lccc}
    \toprule
    Capability & Task 1 & Task 2 & Task 3 \\
    \midrule
    Structured artifact review     & \checkmark & \checkmark & \checkmark \\
    Targeted user intervention     & \checkmark & \checkmark & \checkmark \\
    Pre-execution modification     & \checkmark &            & \checkmark \\
    Improved intent communication  & \checkmark & \checkmark & \checkmark \\
    Preference/style adaptation    &            & \checkmark &            \\
    Remote/cross-device review     &            &            & \checkmark \\
    Terminal-free human oversight  & \checkmark & \checkmark & \checkmark \\
    \bottomrule
  \end{tabular}}
  
  \vspace{-0.2in}
\end{table}

\vspace{-0.15in}
\section{Related Work}
\vspace{-0.05in}
AgentClick relates to three lines of prior work. \textbf{Human-in-the-loop agent systems} study how users can remain involved during agent execution, including through co-planning, action approval, interactive plans, and preference learning~\cite{mozannar2025magenticui,feng2026cocoa,liang2026pahf}. In contrast, AgentClick targets a lighter-weight and more directly deployable review layer for existing terminal-based agents, with explicit preference capture through reason-tagged edits to structured memory files. \textbf{Remote access and terminal interfaces} make terminal-based agents accessible through web or mobile interfaces, allowing users to interact with remotely running agents from personal devices~\cite{siteboon2026cloudcli}. Unlike these systems, AgentClick does not simply expose a chat or terminal view, but provides task-specific interfaces for inspecting, editing, and approving structured artifacts. \textbf{Debugging and execution platforms} support agent development and inspection either by running agents inside dedicated infrastructures or by analyzing their traces after execution~\cite{wang2024openhands,epperson2025agdebugger}. By contrast, AgentClick attaches to existing terminal agents and intervenes before or during execution, enabling structured review rather than post-hoc debugging or sandboxed re-hosting.

\vspace{-0.15in}
\section{Conclusion}
\vspace{-0.05in}
We presented AgentClick, a lightweight browser-based review layer that improves human--agent collaboration for terminal-based agents without modifying the underlying system. By surfacing structured, artifact-specific interfaces for email, planning, code, memory, and trajectory review, AgentClick addresses the visibility, correction, and accessibility gaps of terminal-first interaction. We hope it demonstrates that the review interfaces are a practical path toward more effective and accessible human--agent co-work.

\bibliographystyle{ACM-Reference-Format}
\bibliography{ref}
\clearpage
\section*{Appendix}

\begin{figure}[!htbp]
    \centering
    \begin{subfigure}{0.47\textwidth}
        \centering
        \includegraphics[width=\linewidth]{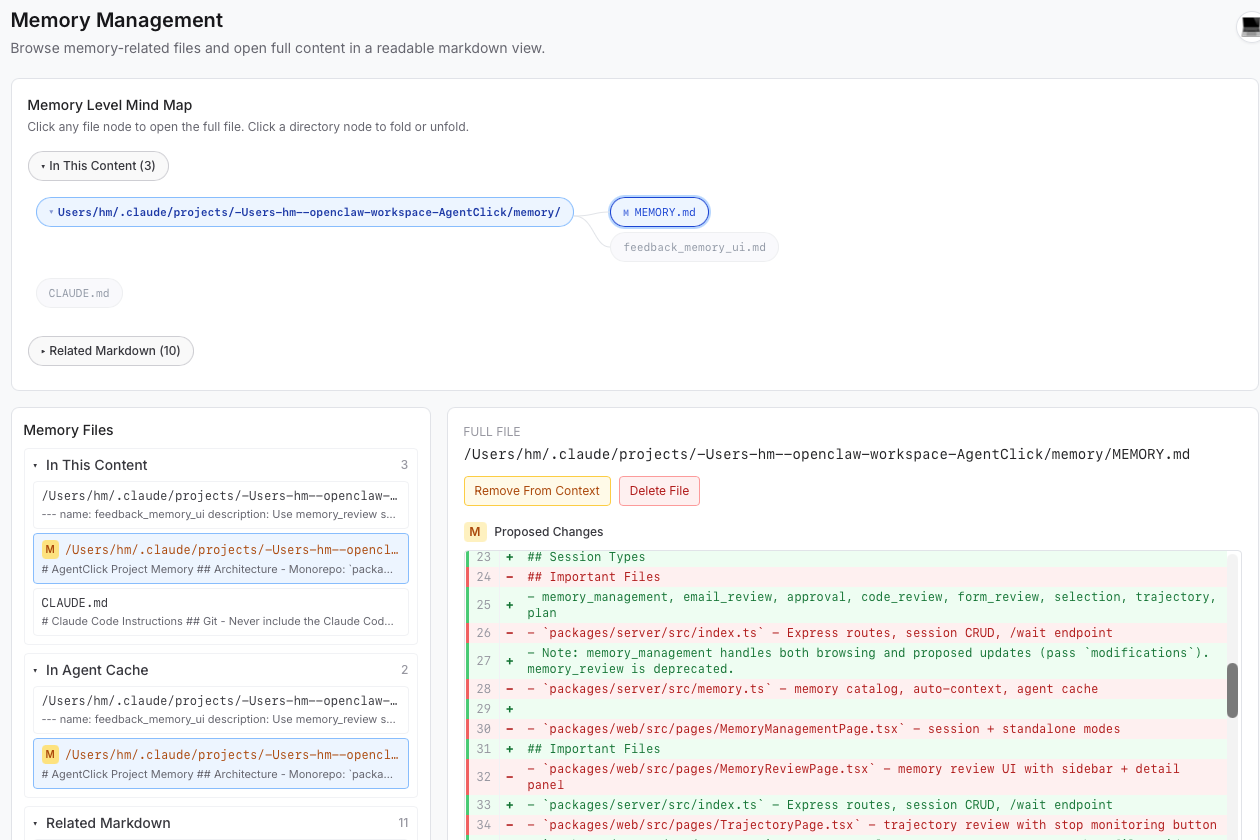}
        \caption{Memory Browser UI}
        \label{fig:sub1}
    \end{subfigure}
    \hfill
    \begin{subfigure}{0.47\textwidth}
        \centering
        \includegraphics[width=\linewidth]{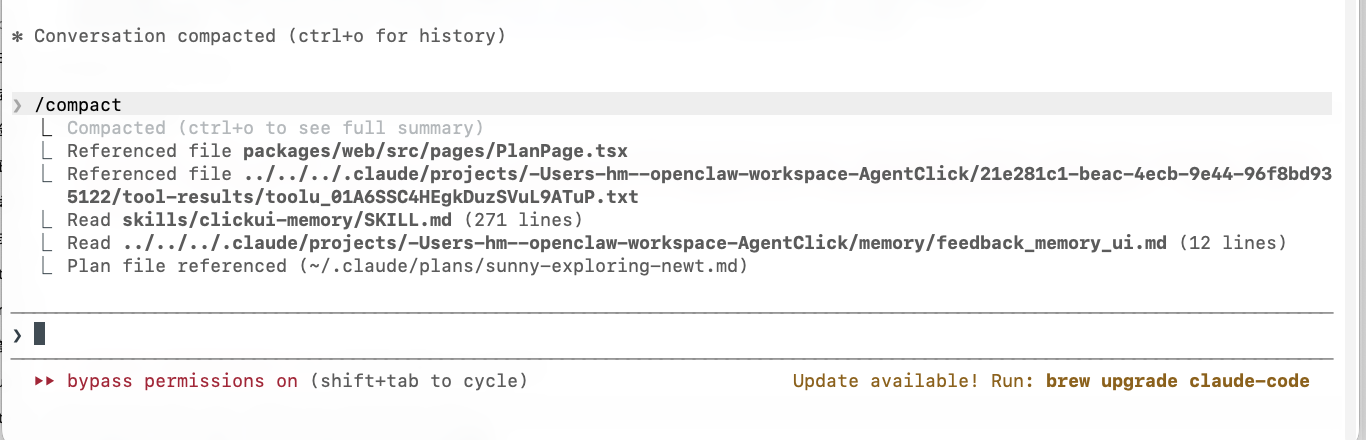}
        \caption{Memory TUI}
        \label{fig:sub2}
    \end{subfigure}
    \caption{\textbf{AgentClick Memory UI.} Before context compaction, the interface surfaces all memory entries modified in the current session alongside a generated session summary. Users can review and edit the summary before it is committed, and selectively choose which entries to load into the next session. }
    \label{fig:memory}
\end{figure}

\begin{figure}[!htbp]
    \centering
    \includegraphics[width=0.95\linewidth]{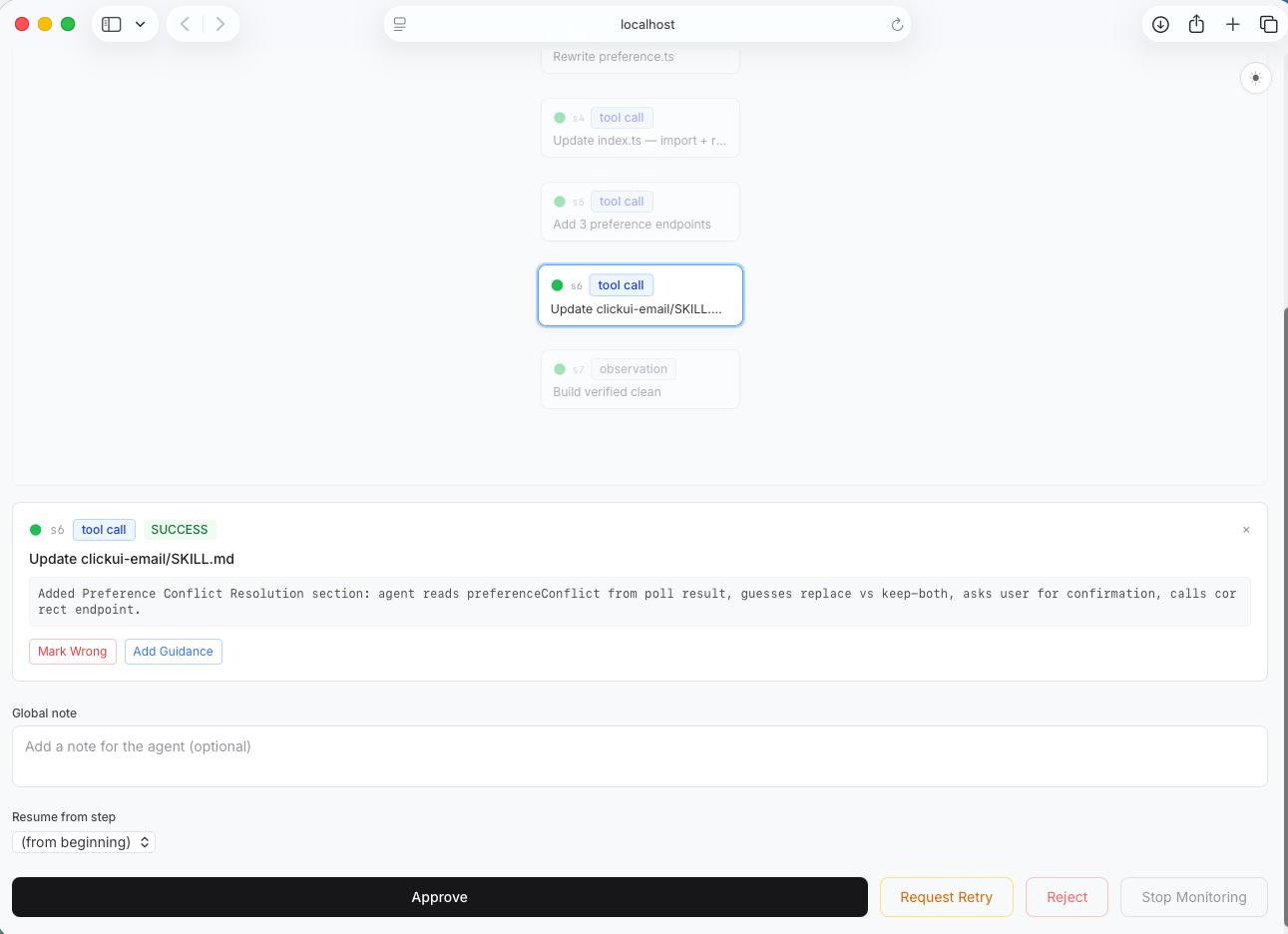}
    \caption{Click UI for trajectory.}
    \label{fig:traj}
\end{figure}
\FloatBarrier

\end{document}